\documentclass[preprint,12pt]{elsarticle}

\usepackage{amssymb}

\journal{Biosystems}

\begin{document}

\begin{frontmatter}



\title{Understanding nature's selection of genetic languages}


\author[label1,label2]{Apoorva D. Patel}

\affiliation[label1]{organization={Centre for High Energy Physics, Indian Institute of Science},
            city={Bengaluru},
            postcode={560012}, 
            country={India}}
\affiliation[label2]{organization={International Centre for Theoretical Sciences},
            city={Bengaluru},
            postcode={560089}, 
            country={India}}

\begin{abstract}
All living organisms use two universal genetic languages in their molecular
biology machinery, one containing four nucleotide bases in its alphabet,
and the other containing twenty amino acids in its alphabet. They can be
understood as the optimal encodings of genetic information for the tasks
they carry out, i.e. replication/transcription for DNA/RNA and translation
for polypeptide chains. These tasks select needed letters of the alphabet
by complementary nucleotide base-pairing, from a collection of molecules
in the cell. The computer science paradigm for this process is database
search; various algorithms for it can be constructed and compared according
to number of attempts (or queries) they need to make to find the correct
nucleotide base-pairing. Grover's search algorithm based on oscillatory
wave dynamics perfectly fits the number of queries needed to search the
genetic alphabets, and it is more efficient than the best Boolean search
algorithm (i.e. binary tree search) that needs a larger number of queries.
This result strongly suggests that the universal genetic languages have
been selected by evolution as the optimal alphabets for the tasks they
carry out, and are not an accident of history. The outstanding challenge
is to demonstrate how Grover's search algorithm would be executed in vivo
by the living organisms.
\end{abstract}




\end{frontmatter}


\section{Biological Facts}

Languages, aperiodic collections of letters of their alphabets, form the
basis of any communication system. In his seminal work \cite{schrodinger1944},
Schr\"odinger argued that such a language passes on the hereditary information
of living organisms. In fact, studies in molecular biology have discovered two
fundamental languages, one consisting of four nucleotide bases that describes
the content of DNA/RNA, and the other consisting of twenty amino acids that
encodes the structure of proteins. Certain general properties of these
languages have been established \cite{watson2014}, which have important
implications regarding how they may have naturally evolved:\\
$\bullet$ The languages of DNA/RNA and proteins are universal. The same 4
nucleotide bases and 20 amino acids are used by all living organisms,
despite the fact that other nucleotide bases and amino acids exist in
living cells. Clearly, evolution has selected specific alphabets.\\
$\bullet$ The information is encoded close to the data compression limit
for genes (akin to zipping of computer data files) and maximal packing for
proteins (which reduces volume). Such compactifications remove unnecessary
correlations and indicate that the information storage has been optimised.\\
$\bullet$ Mutations occur during evolution. Darwinian selection can
eliminate suboptimal competitors, provided there is sufficient time to
explore various competing possibilities. That can make the surviving
candidates universal, instead of their appearance as an accidental event.

Taking note of these properties, we argue that the universal genetic languages
are what they are because they have evolved to be optimal for the tasks they
carry out, in a competition among various possibilities. The tasks carried
out by genetic languages are one-dimensional sequential assembly of letters
in chains \cite{watson2014}. Complementary building blocks identified by
nucleotide base-pairing are added on top of pre-existing master templates,
proceeding one letter at a time in case of replication and transcription for
DNA/mRNA, and proceeding three letters at a time in case of translation from
mRNA to polypeptide chains. This assembly process is simple enough to
explicitly determine its best form in various algorithmic settings. The
complementary base-pairing with the master template is a binary step, either
it happens or it does not. The correct candidate for base-pairing is picked
up from the collection of molecules in the cell by a search process; familiar
examples of such a database search are how we find a word in a dictionary
or how we locate a webpage on internet. In the computer science framework,
with distinguishable building blocks, the base-pairing is a binary search
oracle query that checks for a particular property of the desired object,
and is answered by a readymade look-up table. The algorithm optimisation
problem is then to find the suitable complementary building blocks (i.e.
the nucleotide base or the amino acid), using as few binary oracle queries
as possible in the database search process.

\section{Optimal Search Algorithm}

The problem to be tackled is the assembly of a growing chain using objects
selected from a database using binary oracle queries. The best Boolean
algorithm for it is the binary tree search \cite{knuth1998vol3}, exemplified
by how we look up a word in a dictionary by repeatedly checking whether the
desired word is before or after the present location (after starting the
search somewhere in the middle of the dictionary). For a database of size
$N$, it takes $\log_2N$ binary oracle queries. One complementary pair of
nucleotide bases is sufficient to encode this algorithm. The algorithm
also requires the database to be sorted (as in case of a dictionary),
but we may assume that nature would have found a way to do that were this
algorithm advantageous. In reality, nature has selected more complicated
genetic languages containing four nucleotide bases instead of two.

Lov Grover discovered a different search algorithm that asymptotically
requires $Q=O(\sqrt{N})$ binary oracle queries, with an unsorted database
\cite{grover1996}. Putting aside all considerations of large-$N$ behaviour,
it is striking to observe that $\sqrt{N}\le\log_2N$ for $4 \le N \le 16$.
It means that Grover's search algorithm is superior to the best Boolean
search algorithm, in the range of $N$ relevant to search of letters in
genetic alphabets.

The relevant range for superiority of Grover's search algorithm expands
when more accurate solutions for the algorithm are considered. The leading
asymptotic behaviour reduces the number of binary oracle queries to
$\frac{\pi}{4}\sqrt{N}$, while the exact solution reduces it even more,
as per
\begin{equation}
\label{querysoln}
(2Q+1)\sin^{-1}\frac{1}{\sqrt{N}} = \frac{\pi}{2} ~.
\end{equation}
It means that Grover's search algorithm can have a small failure rate and
still remain superior to binary tree search, in the relevant range of $N$.

It should be noted that Grover's algorithm has been shown to be optimal
for all values of $N$, using variational methods \cite{zalka1999}.
Moreover, $Q=1$ for $N=4$ is an exact solution of Eq.(\ref{querysoln})
that maps to identification of the four nucleotide bases of DNA/RNA with
a single base-pairing, while $Q=3$ for $N=20$ is an approximate solution
(with a tiny error) of Eq.(\ref{querysoln}) that maps to identification
of the twenty amino acids in polypeptide chains with three base-pairings.

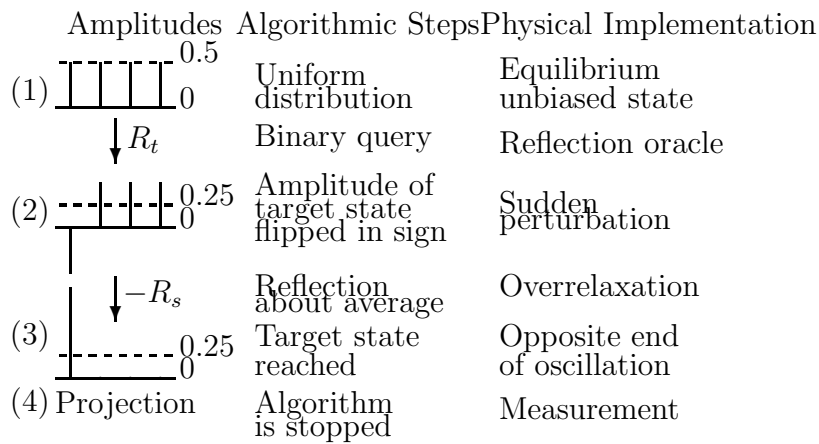
\begin{figure}[t]
\setlength{\unitlength}{0.5mm}
\hspace{5mm}
\begin{picture}(250,120)
  \thicklines
\put(15,105){\makebox(0,0)[bl]{Amplitudes}}
\put(60,105){\makebox(0,0)[bl]{Algorithmic Steps}}
\put(125,105){\makebox(0,0)[bl]{Physical Implementation}}
  \put( 0,87){\makebox(0,0)[bl]{(1)}}
  \put(12,87){\line(1,0){32}}
\put(13,99){\line(1,0){2}} \put(17,99){\line(1,0){2}} \put(21,99){\line(1,0){2}}
\put(25,99){\line(1,0){2}} \put(29,99){\line(1,0){2}} \put(33,99){\line(1,0){2}}
\put(37,99){\line(1,0){2}} \put(41,99){\line(1,0){2}}
  \put(45,87){\makebox(0,0)[bl]{0}} \put(45,99){\makebox(0,0)[bl]{0.5}}
  \put(16,87){\line(0,1){12}} \put(24,87){\line(0,1){12}}
  \put(32,87){\line(0,1){12}} \put(40,87){\line(0,1){12}}
  \put(65,93){\makebox(0,0)[bl]{Uniform}}
  \put(65,87){\makebox(0,0)[bl]{distribution}}
  \put(130,93){\makebox(0,0)[bl]{Equilibrium}}
  \put(130,87){\makebox(0,0)[bl]{unbiased state}}
  \put(28,84){\vector(0,-1){12}}
  \put(31,75){\makebox(0,0)[bl]{$Q_t$}}
  \put(65,75){\makebox(0,0)[bl]{Binary query}}
  \put(130,75){\makebox(0,0)[bl]{Reflection oracle}}
  \put( 0,55){\makebox(0,0)[bl]{(2)}}
  \put(12,55){\line(1,0){32}}
\put(13,61){\line(1,0){2}} \put(17,61){\line(1,0){2}} \put(21,61){\line(1,0){2}}
\put(25,61){\line(1,0){2}} \put(29,61){\line(1,0){2}} \put(33,61){\line(1,0){2}}
\put(37,61){\line(1,0){2}} \put(41,61){\line(1,0){2}}
  \put(45,55){\makebox(0,0)[bl]{0}} \put(45,61){\makebox(0,0)[bl]{0.25}}
  \put(16,55){\line(0,-1){12}} \put(24,55){\line(0,1){12}}
  \put(32,55){\line(0,1){12}} \put(40,55){\line(0,1){12}}
  \put(65,62){\makebox(0,0)[bl]{Amplitude of}}
  \put(65,56){\makebox(0,0)[bl]{target state}}
  \put(65,50){\makebox(0,0)[bl]{flipped in sign}}
  \put(130,59){\makebox(0,0)[bl]{Sudden}}
  \put(130,53){\makebox(0,0)[bl]{perturbation}}
  \put(28,42){\vector(0,-1){12}}
  \put(30,34){\makebox(0,0)[bl]{$R_s$}}
  \put(65,37){\makebox(0,0)[bl]{Reflection}}
  \put(65,31){\makebox(0,0)[bl]{about average}}
  \put(130,37){\makebox(0,0)[bl]{Overrelaxation}}
  \put( 0,22){\makebox(0,0)[bl]{(3)}}
  \put(12,15){\line(1,0){32}}
\put(13,21){\line(1,0){2}} \put(17,21){\line(1,0){2}} \put(21,21){\line(1,0){2}}
\put(25,21){\line(1,0){2}} \put(29,21){\line(1,0){2}} \put(33,21){\line(1,0){2}}
\put(37,21){\line(1,0){2}} \put(41,21){\line(1,0){2}}
  \put(45,15){\makebox(0,0)[bl]{0}} \put(45,21){\makebox(0,0)[bl]{0.25}}
  \put(16,15){\line(0,1){24}}
  \put(24,15){\circle*{1}} \put(32,15){\circle*{1}} \put(40,15){\circle*{1}}
  \put(65,22){\makebox(0,0)[bl]{Target state}}
  \put(65,16){\makebox(0,0)[bl]{reached}}
  \put(130,22){\makebox(0,0)[bl]{Opposite end}}
  \put(130,16){\makebox(0,0)[bl]{of oscillation}}
  \put( 0,4){\makebox(0,0)[bl]{(4)}}
  \put(12,4){\makebox(0,0)[bl]{Observation}}
  \put(65,4){\makebox(0,0)[bl]{Algorithm}}
  \put(65,-2){\makebox(0,0)[bl]{is stopped}}
  \put(130,4){\makebox(0,0)[bl]{Measurement}}
  \put(200,60){\makebox(0,0)[bl]{Evolution is:}}
  \put(200,50){\makebox(0,0)[bl]{$|t\rangle=(R_sQ_t)^Q|s\rangle$}}
\end{picture}
\caption{The steps of Grover's search algorithm for the simplest case of
searching four items in the database, when the first item is marked by the
binary oracle query. The left column depicts the amplitudes of the four
states that evolve coherently, with the dashed lines showing their average
values. The middle column describes the algorithmic steps, and the right
column mentions their physical implementation. This search algorithm is
superior to the binary tree seach by a factor of two. The algorithmic
evolution in case of multiple binary oracle queries is expressed by the
equation on the extreme right, showing evolution of the initial state
$|s\rangle$ to the target state $|t\rangle$ with the standard Dirac notation.
\label{onefromfour}}
\end{figure}

\section{Physical Implementations of Grover's Algorithm}

The steps of Grover's search algorithm are illustrated in Figure 1 for the
simplest instance, $N=4$ and $Q=1$. The algorithm begins with an unbiased
uniform superposition state, and applies two reflection operators alternately
for $Q$ iterations, until the target state is reached. One of the reflection
operators is the response to the binary oracle query, and the other is the
reflection across the uniform superposition state. A specific criterion is
needed to stop the iterative steps, when the target state is reached.

The key feature of the algorithm is oscillatory wave dynamics that allows
superposition of its modes. Once coherent wave modes are available, i.e.
relative phases of the oscillating components are maintained, the algorithm
needs nothing more than the two reflection operations. Figure 1 depicts how
the algorithm unambiguously identifies one out of four items in the database
using a single binary oracle query. In contrast, the binary tree search would
only identify one out of two items in the database with one binary oracle
query.

Figure 1 also points out the physical steps required to implement Grover's
search algorithm, and they are quite simple. It suffices to have coherent
wave modes that can be superposed and phase-shifted, and so classical wave
dynamics can implement the algorithm as well \cite{grover2002,patel2006}.
The phase-shift required to perturb synchronised oscillations can be
generated in many ways, such as scattering from an impurity, formation
of chemical bonds and constraints imposed by boundaries.

In general, Grover's search algorithm amplifies the amplitude of the target
state, while suppressing the amplitudes of the non-target states. It is
referred to as a search algorithm due to the quantum interpretation of
$|{\rm amplitude}|^2$ as probability. In the classical wave dynamics setting,
however, $|{\rm amplitude}|^2$ represents energy of the wave mode. Moreover,
in a coupled system of oscillators, the centre-of-mass mode plays the role
of the unbiased uniform superposition state, and reflection operations can
be implemented as elastic collisions. Such a scenario has been demonstrated
in a high school science project \cite{patel2007a}. Furthermore, energy
amplification of the target mode beyond a threshold can initiate a new
reaction that effectively terminates the algorithm. The consequences can
be dramatic in processes whose rates are governed by the Boltzmann factor,
$\exp(-E/kT)$, where the energy appears in the exponent.

A point to note is that both the classical wave version and the quantum
version of Grover's algorithm have identical oracle query count, but the
classical version needs $N$ distinct wave modes while the quantum one
requires $\log_2N$ qubits. The classical wave version therefore needs
more spatial resources for the algorithm than the quantum one, although
the temporal resources remain the same.

\section{Grover's Algorithm in Natural Setting}

At the molecular level, quantum coherence is hard to sustain in biological
systems, but features appropriate for implementing Grover's search algorithm
in the classical wave version exist:\\
$\bullet$ Various biomolecules are obtained by digesting food, and then
reassembled according to specific prescriptions. In this metabolic process,
the elementary components are randomly floating around in the cellular
environment, and their assembly in a specific order is unsorted database
search. All candidate molecules are made available by a diffusion dominated
transport, and that suffices for the execution of the search algorithm.\\
$\bullet$ In the biological context, time is highly precious while space is
fairly expendable. A small difference in population growth rates, even an
advantage of a fraction of a percent, is sufficient for one species to
overwhelm another over many generations. The number of seeds produced by
a plant, and the number of cells in a tissue, are examples of large scale
parallelisation of the effort required for carrying out the desired task.
Thus the extra cost of spatial resources for the classical wave search
can indeed be tolerable for small values of $N$, and the binary tree search
can be beaten in the race for temporal advantage.\\
$\bullet$ Tasks involving efficient transfer of energy, from one form to
another, have high priority in the non-equilibrium metabolic processes of
living organisms. The kinetic isotope effect demonstrates that the rates of
chemical processes are influenced by the vibrational motion of the atoms
involved \cite{westheimer1961}, in addition to their electronic energies.
Various isotopes of an element have the same electronic configuration but
different nuclear masses, so the vibrational energy of their nuclei differ
while the electronic energy remains the same. Isotope substitution therefore
allows detection of coupling between vibrational and electronic degrees of
freedom through changes in chemical reaction rates, and such changes have
been observed in enzyme catalysis \cite{huelga2013}. The effect is the
largest when deuterium is substituted for hydrogen (say in a hydrogen bond),
where the ratio of nuclear masses is the largest.

Beyond these features, it is worth noting that:\\
(a) Grover's algorithm is quite robust against perturbations. Due to
its optimal nature, its outcome is affected only at the second order in
physical disturbances. (The first derivative of a smooth function vanishes
at its minimum.)\\
(b) $Q=2$ for $N=10$ is an approximate solution of Eq.(\ref{querysoln}).
It agrees with the partition of the amino acids, into two classes of ten
each, by a doublet genetic code that is likely to have preceded the current
triplet genetic code during evolution \cite{patel2005,wu2005,rodin2006}.\\
(c) For successful search, Grover's algorithm requires the coherent
superposition of wave modes to survive till the execution of algorithm
completes. But an apparent superposition would suffice, where the molecular
diffusion cycling time between different selection options is short compared
to the overall nucleotide base-pairing time.\\
(d) The nucleotide base-pairing takes place via hydrogen bonds, which are
highly sensitive to vibrational motion.

Replacement of the fragile quantum version of Grover's algorithm by its
stable classical wave version keeps its binary oracle query advantage
unchanged. Given the simplicity of the algorithm (the required steps are
only those illustrated in Figure 1), it is perfectly plausible that
biological evolution discovered it by trial-and-error exploration, and
thereafter its optimality made it an inalienable part of life.

\section{Experimental Verification}

Preceding sections have described how Grover's search algorithm is the
optimal solution for the genetic languages to implement their tasks, and
how its simple steps can be mapped to the physical processes existing in
living cells. That makes a compelling case that the classical wave version
is indeed the reason behind the observed universal form of the genetic
languages. Several scenarios have been proposed for what the details of
the map could be \cite{patel2007}, but the ultimate test must come from
experimental verification. Two possibilities come to mind:\\
(1) An indirect, and easier to implement, test would check the optimality
of the algorithm by a competition between the natural genetic languages and
their artificially constructed competitors \cite{patel2001a}. The competitors
would use less or more number of letters in their genetic alphabet, while
encoding the same amount of genetic information. The winner would be the
language that carries out its required chain assembly faster. Technology
for performing such tests is available, though competition tests with the
existing polymerase support system tuned to the natural genetic languages
would leave the fairness of comparison in doubt.\\
(2) A direct test would be to observe the intermediate steps of the genetic
replication/transcription and translation processes. Such a check of dynamics
of the algorithm is difficult, and not yet possible in vivo, but the progress
in high-speed photography that observes intermediate steps of chemical
reactions should take us there some day.

Such a confirmation would tell us that indeed the advantage of classical
wave search made nature select the universal genetic languages observed
today, compared to their minimal binary versions. Starting from scratch,
forming a binary language consisting of a single pair of complementary
nucleotide bases could have been the first step in molecular biology, but
nature evidently found reason enough to go beyond that and evolve more
complicated genetic languages.




\end{document}